\documentclass[sort&compress,final,5p,twocolumn,times,fixfloat]{elsarticle}

\journal{Physics Letters B}

\usepackage{lmodern}
\usepackage{lineno}
\usepackage{lipsum}
\usepackage{url}
\usepackage{geometry}
\usepackage{graphicx}
\usepackage{amsmath}
\usepackage{amssymb}
\usepackage{latexsym}
\usepackage{amsfonts}
\usepackage[hidelinks]{hyperref}
\usepackage{mathtools}
\usepackage{multirow}
\usepackage{sidecap}
\usepackage{soul}
\usepackage{braket}
\usepackage{slashed}
\usepackage{booktabs}
\usepackage{bbm}
\usepackage{bm}
\usepackage{scalerel}
\usepackage{tikz}
\usetikzlibrary{svg.path}

\def\nostrocostruttino#1\over#2{\mathrel{\mathop{\kern 0pt \rlap
{\hbox{$#1$}}} \hbox{\kern-.125em $#2$}}}

\newcommand{\be}{\begin{equation}}
\newcommand{\ee}{\end{equation}}
\newcommand{\bea}{\begin{eqnarray}}
\newcommand{\eea}{\end{eqnarray}}

\usepackage{xcolor}
\DeclareUnicodeCharacter{2212}{-}

\begin{document}

\title{Transverse $\Lambda$ polarization in unpolarized $pp\rightarrow \, {\rm jet}\, \Lambda^\uparrow\, X $ }

\date{\today}

\author[add1,add2]{Umberto D'Alesio
}
\ead{umberto.dalesio@ca.infn.it}

\author[add3]{Leonard Gamberg
}\ead{lpg10@psu.edu}

\author[add2]{Francesco Murgia
}
\ead{francesco.murgia@ca.infn.it}

\author[add4]{Marco Zaccheddu
}
\ead{zacch@jlab.org}

\address[add1]{Dipartimento di Fisica, Universit\`a di Cagliari, Cittadella Universitaria, I-09042 Monserrato (CA), Italy}
\address[add2]{INFN, Sezione di Cagliari, Cittadella Universitaria, I-09042 Monserrato (CA), Italy}
\address[add3]{Division of Science, Penn State Berks, Reading, PA 19610, USA}
\address[add4]{Theory Center, Jefferson Lab, 12000 Jefferson Avenue, Newport News, Virginia 23606, USA}

\begin{abstract}

In this Letter, we investigate the spontaneous transverse polarization of $\Lambda$ hyperons produced in unpolarized $pp$ collisions inside a jet, by adopting a TMD approach where transverse momentum effects are included only in the fragmentation process. We will present predictions based on the parametrizations of the $\Lambda$ polarizing fragmentation function as extracted from fits to Belle $e^+e^-$ data. These estimates will be compared against preliminary STAR data. We will then be able to explore the universality properties of the quark polarizing fragmentation function and, for the first time, the role of its gluon counterpart.

\end{abstract}

\begin{keyword}
 hadron-in-jet  \sep $\Lambda$ polarization \sep TMD approach \sep polarizing fragmentation function  - JLAB-THY-24-3993
\end{keyword}

\maketitle

\section{Introduction}
\label{1_sIntro}

Understanding the  hadronization  of partons in terms of transverse spin  and their  correlations with intrinsic transverse momentum degrees of freedom remains an outstanding challenge in unfolding the partonic structure of hadrons.
In this context, one of the most fundamental problems is to reveal the dynamical mechanism that provides the spontaneous transverse polarization of $\Lambda$ fragmentation in unpolarized lepton-lepton, lepton-hadron and hadron-hadron scattering within the field theory of partonic interactions, quantum chromodynamics (QCD).

QCD provides the theoretical framework to study the partonic correlations of hadron structure in conjunction with transverse momentum dependent (TMD) factorization theorems~\cite{Collins:1981uk,Collins:1981va, Collins:1984kg,collins_2011}.
TMD factorization provides a framework that links perturbative parton dynamics of the quark and gluon structure to the rich nonperturbative three dimensional (3-D) momentum  structure of hadrons~\cite{Boussarie:2023izj}.
It is characterized by the presence of two ordered energy scales: a small one (e.g.~the transverse momentum unbalance of the two hadrons produced in opposite hemispheres in $e^+e^-$ processes or the transverse momentum of the final hadron in lepton-hadron semi-inclusive deep-inelastic scattering (SIDIS))\footnote{Here we consider only hadron-production processes.} and a large one (e.g. the virtuality of the exchanged photon). A fundamental prediction of TMD factorization for this class of processes is that the nonperturbative intrinsic structure in the fragmentation process is universal~\cite{Metz:2002iz,Boer:2003cm,Collins:2004nx,Yuan:2009dw,Boer:2010ya}. Universality and scale evolution are  essential properties that allow one to study hadron structure in different processes and across a wide range of energies.
Moreover, in this context, the so-called naive time reversal odd (T-odd), transverse momentum dependent fragmentation functions (TMDFFs), like the Collins function~\cite{Collins:1992kk} and the polarizing fragmentation function (polFF)~\cite{Mulders:1995dh, Anselmino:2000vs} are processes independent~\cite{Metz:2002iz,Boer:2003cm,Collins:2004nx,Yuan:2007nd,Meissner:2008yf,Gamberg:2008yt,Yuan:2009dw,Gamberg:2010uw,Boer:2010ya}. This is to be compared with the modified T-odd universality in the initial state for the Sivers~\cite{Sivers:1989cc} and Boer-Mulders~\cite{Boer:1999mm} functions~\cite{Collins:2002kn,Brodsky:2002rv,Boer:2003cm}.

Early phenomenological studies of the $\Lambda$ polarizing fragmentation function in unpolarized proton-proton collisions and SIDIS processes were carried out in Refs.~\cite{Anselmino:2000vs,Anselmino:2001js}.
More recently, experimental data collected by the Belle Collaboration~\cite{Belle:2018ttu} for the transverse $\Lambda,\bar\Lambda$ polarization in almost back-to-back two-hadron production in $e^+e^-$ processes has resulted in new phenomenological analyses.
Studies within a TMD scheme at fixed scale were presented in Refs.~\cite{DAlesio:2020wjq,Callos:2020qtu}, while subsequent extractions, implementing the Collins-Soper-Sterman (CSS) motivated TMD evolution~\cite{Collins:1981uk,Collins:1981va, Collins:1984kg}, were carried out in Refs.~\cite{Gamberg:2021iat,Kang:2021kpt,Li:2020oto,Chen:2021hdn, DAlesio:2022brl}. Then, in Ref.~\cite{DAlesio:2023ozw} the role of ${SU}(2)$ symmetry (see also Refs.~\cite{Chen:2021hdn, Chen:2021zrr}) as well as of the charm contribution was explored with some detail.

In recent years the study of the transverse momentum distribution of hadrons inside jets has garnered much attention  as a tool to explore the hadronization mechanism~\cite{Yuan:2007nd,DAlesio:2010sag,Kang:2017btw,Kang:2017glf}. As  this pertains to studying TMD fragmentation, these processes complement the benchmark ones, SIDIS and single and double semi-inclusive hadron production in electron-positron annihilation.
A significant appeal to studying single-inclusive hadron production within a jet in $pp$ collisions is due to the fact that one can employ the collinear parton distribution functions (PDFs), which allows for a direct probe of the
transverse momentum dependent hadronization process.
Theoretical developments on hadron in jet factorization theorems given in terms of transverse momentum dependent jet-fragmentation functions (TMDJFFs) were presented in Refs.~\cite{Bain:2016rrv,Kang:2017glf,Kang:2020xyq}, where it is demonstrated that these TMDJFFs are directly related to the ordinary TMDFFs when the transverse momentum of the hadron is measured with respect to the standard jet axis. A comprehensive theoretical analysis for the distribution of polarized hadrons within jets in electron-proton and proton-proton collisions was performed in Refs.~\cite{Kang:2020xyq,Kang:2023elg,Kang:2021ffh}.

For the production of a transversely polarized spin-1/2 hadron, the polFF couples directly to the collinear unpolarized PDFs in the initial state. With accurately determined PDFs, one can directly probe the self-analyzing fragmentation mechanism, in principle allowing for a further determination of the $\Lambda$ hyperon polarizing TMDFF, through the measurement of its transverse polarization. 
Moreover, this process can eventually serve as an additional test of the universality of T-odd fragmentation functions.

In this context, first studies of the universality of the Collins effect in hadron in jet processes were performed in Refs.~\cite{Yuan:2007nd,DAlesio:2010sag}.
More recent phenomenological studies of hadron in jet Collins azimuthal asymmetries were carried out incorporating evolution effects~\cite{Kang:2017btw}, and also in  the generalized parton model~\cite{DAlesio:2017bvu}\footnote{These analyses yielded similar results.}.

Quite recently a new opportunity has presented itself with the availability of preliminary data on transverse polarization of $\Lambda$'s produced inside a jet in unpolarized proton-proton collisions at RHIC from the STAR Collaboration~\cite{Gao:2024dxl,spinGao}. This measurement in principle can provide further constraints on the polFFs and might eventually be used in global analyses.

In this letter, we will carry out a preliminary phenomenological study using our recent extractions of the polFFs from fits to $e^+e^-$ annihilation processes~\cite{DAlesio:2022brl,DAlesio:2023ozw}, to investigate the spontaneous transverse polarization of $\Lambda$ hyperons produced in unpolarized proton-proton collisions inside a jet.
Our analysis can be considered as a first attempt to check the predicted universality properties of the polFFs, a fundamental issue as mentioned above.
Another important aspect, never treated before, is that for this  class  of processes, by contrast with $e^+e^-$ and SIDIS, one can directly access gluon TMDFFs, since all partons enter at the same perturbative order. This would open a window on the study of the still unknown polarizing fragmentation function for gluons, as we will discuss.

The letter is organized as follows: after reviewing the main aspects of the formalism and all basic formulas in Section~\ref{2_formalism}, we present the phenomenological analysis, and  then our theoretical predictions against STAR data in Section~\ref{4_phenomenology}. Our conclusions and final remarks are collected in Section~\ref{5_conclusions}.

\section{Transverse $\Lambda$ polarization in unpolarized $pp$ collisions}
\label{2_formalism}

In this section, we provide the main formulas to compute the transverse polarization, $P_T^\Lambda$, of $\Lambda(\bar\Lambda)$ hyperons (or any spin-1/2 hadron) produced within a jet in unpolarized hadron-hadron ($AB$) collisions,
\begin{equation}
A(p_A)\,B(p_B)\rightarrow {\rm jet}(p_{\rm{j}})\,  \Lambda^\uparrow(p_\Lambda)\,X\, ,
\end{equation}
where $p_A,p_B, p_j, p_\Lambda$ are the four-momenta of the incoming hadrons, the jet, and the produced $\Lambda$ respectively.
This observable is defined as
\be
\label{polT}
P_T^\Lambda(\bm{p}_{\rm j}, \xi,\bm{p}_{\perp\Lambda} ) = \frac{d\sigma^\uparrow - d\sigma^\downarrow}{d\sigma^\uparrow+ d\sigma^\downarrow} = \frac{d\Delta\sigma}{d\sigma_{\rm unp}}\,,
\ee
where the differential cross section for the production of a $\Lambda$ within a jet transversely polarized with respect to the $\Lambda$-jet plane is
\be
d\sigma^{\uparrow(\downarrow)} \equiv E_{\rm j}\, \frac{d\sigma^{AB\to {\rm jet} \,\Lambda^{\uparrow(\downarrow)} X}}{d^3\bm{p}_{\rm j}d\xi d^2\bm{p}_{\perp\Lambda}} \,,
\ee
and $d\sigma_{\rm unp}$ is the unpolarized cross section, while $\xi$ is the $\Lambda$ light-cone momentum fraction and $p_{\perp\Lambda}\equiv|\bm{p}_{\perp\Lambda}|$ its transverse momentum with respect to the fragmenting jet.

We will employ a leading order (LO) factorization framework, with a collinear configuration for the initial state followed by TMD factorization for the final state, similarly to the scheme adopted in Refs.~\cite{Yuan:2007nd,Kang:2017btw}.
More precisely, the full differential cross section for the production of a transversely polarized $\Lambda$ within a jet can be expressed as:
\be
\label{pplup}
\begin{aligned}
 E_{\rm j}\, \frac{d\sigma^{AB\to {\rm jet} \Lambda^\uparrow X}}{d^3\bm{p}_{\rm j}d\xi d^2\bm{p}_{\perp\Lambda}}
=  \sum_{a,b,c,d}\int dx_adx_b\, \frac{\alpha_s^2}{\hat s}f_{a/A}(x_a) f_{b/B}(x_b) \\
\times\, |\overline{M}_{ab\to cd}|^2\delta(\hat s + \hat t +\hat u) \,\hat D_{\Lambda^\uparrow/c}(\xi,\bm{p}_{\perp\Lambda})\,,
\end{aligned}
\ee
where $f_{a,b}(x)$'s are the collinear PDFs, $\hat D_{\Lambda^\uparrow/c}$ is the TMDFF for an unpolarized parton $c$ fragmenting into a transversely polarized $\Lambda$, and $\hat s$, $\hat t$ and $\hat u$ are the standard partonic Maldelstam invariants. Lastly,
$|\overline{M}_{ab\to cd}|^2$ are the amplitudes squared for the hard elementary process $ab\to cd$, averaged(summed) over initial(final) spins and colors.
They are normalized so that the unpolarized partonic cross, for a collinear collision, is given by
\be
\frac{d\sigma^{ab\to cd}}{d\hat t} = \frac{1}{16\pi \hat s^2} \,|\overline{M}|^2\, ,
\ee
where here and in the following we  drop the partonic subscripts on the hard scattering  amplitudes.

In a LO pQCD approach the scattered parton $c$ in the hard elementary process $ab \to cd$ is identified with the observed fragmenting jet: $c\equiv $ jet. Notice that the scale dependence of the nonperturbative  functions, i.e.~PDFs and TMDFFs (even if different in nature), has been understood in Eq.~(\ref{pplup}) and will be properly taken into account in the following analysis.

Let us summarize briefly the kinematics adopted.  We will work in the $AB$ center-of-mass (cm) frame, with $AB$ along the $z$ axis and the jet laying in the $xz$ plane, keeping hadron mass effects only in the final state. As discussed in Refs.~\cite{DAlesio:2020wjq, DAlesio:2022brl, DAlesio:2023ozw} 
these could eventually play a role, at least in some kinematical regions.

The four-momenta (for massless initial hadrons and  partons) are given as
\be
\begin{aligned}
&p_A^\mu = \frac{\sqrt s}{2}(1,0,0,1), \quad p_B^\mu  = \frac{\sqrt s}{2}(1,0,0,-1),\\
&
p_\Lambda^\mu  = (E_\Lambda, \bm{p}_\Lambda),\;{\rm with}\\
&
\bm{p}_\Lambda = |\bm{p}_\Lambda| (\sin\theta_\Lambda\cos\phi_\Lambda, \sin\theta_\Lambda\sin\phi_\Lambda,\cos\theta_\Lambda),\\
&p_a^\mu  =x_a \frac{\sqrt s}{2}(1,0,0,1), \quad p_b^\mu  =x_b \frac{\sqrt s}{2}(1,0,0,-1),  \\
&p_{\rm j}^\mu = E_{\rm j} (1, \sin\theta_{\rm j}, 0,\cos\theta_{\rm j}) = p_{\rm{j}T} (\cosh\eta_{\rm j},1,0,\sinh\eta_{\rm j})\,,
\end{aligned}
\ee
where $s$ is the cm energy squared, $\eta_{\rm j}=-\log[\tan (\theta_{\rm j}/2)]$, is the jet pseudorapidity and $p_{\rm{j}T}\equiv |\bm{p}_{\rm{j}T}|$ its transverse momentum in the $AB$ cm frame.
Moreover, in the jet helicity frame the $\Lambda$ four-momentum can be expressed as
\be
\begin{aligned}
p_\Lambda^\mu  =& \Big(\xi E_{\rm j}\Big(
1+\frac{p_{\perp\Lambda}^2+m_\Lambda^2}{4\xi^2E_{\rm j}^2}\Big),p_{\perp\Lambda}\cos\tilde\phi_\Lambda,\\
& \hspace*{1.0cm} p_{\perp\Lambda}\sin\tilde\phi_\Lambda,
\xi E_{\rm j} \Big(
1-\frac{p_{\perp\Lambda}^2+m_\Lambda^2}{4\xi^2E_{\rm j}^2}\Big)\Big) \,,
\end{aligned}
\ee
where $\tilde\phi_\Lambda$ is the $\Lambda$ azimuthal angle around the jet (parton) direction of motion. The partonic Mandelstam invariants are
\be
\begin{aligned}
\hat s = x_ax_b s, \quad \hat t&= -x_a\sqrt s E_{\rm j} (1-\cos\theta_{\rm j}), \\
\hat u &=-x_b\sqrt s E_{\rm j} (1+\cos\theta_{\rm j})\,.
\end{aligned}
\ee

The numerator of Eq.~(\ref{polT}), $d\Delta\sigma$, involves the polarizing fragmentation function, giving the probability that an unpolarized parton fragments into a transversely polarized
spin-1/2 hadron (a $\Lambda$ hyperon in the present study). This is defined as
\be
\begin{aligned}
\label{polarizing}
\Delta \hat D_{\Lambda^\uparrow/c} (\xi,\bm{p}_{\perp\Lambda}) = \hat D_{\Lambda^\uparrow/c}(\xi,\bm{p}_{\perp\Lambda}) - \hat D_{\Lambda^\downarrow/c}(\xi,\bm{p}_{\perp\Lambda}) \\
= \Delta D_{\Lambda^\uparrow/c} (\xi,p_{\perp\Lambda})\, \hat{\bm{P}}^\Lambda \cdot (\hat{\bm{p}}_c\times \hat{\bm{p}}_{\perp\Lambda})\,,
\end{aligned}
\ee
where $\hat{\bm{P}}^\Lambda$ is the $\Lambda$ spin-polarization vector and $\hat{\bm{p}}_c$, $\hat{\bm{p}}_{\perp\Lambda}$ are unit vectors.
To better clarify the above expressions, we recall that according to our “hat-convention” the quantities like $\hat D$ (or $\Delta\hat D$) depend on $\bm{p}_\perp$, including its phase, while quantities like $D$ (or $\Delta D$) do not depend on phases anymore, as such dependence has been explicitly factorized out.
Another common notation adopted in the literature~\cite{Bacchetta:2004jz} is
\be
\Delta D_{\Lambda^\uparrow/c} (\xi,p_{\perp\Lambda}) = \frac{p_{\perp\Lambda}}{\xi m_\Lambda} D_{1T}^{\perp c}(\xi,p_{\perp\Lambda})\,.
\ee
We also recall that the cross product entering the definition of the polFF in Eq.~(\ref{polarizing}) can be expressed as follows
\be
\hat{\bm{P}}^\Lambda \cdot (\hat{\bm{p}}_c\times \hat{\bm{p}}_{\perp\Lambda})
= \sin(\tilde\phi_{S_\Lambda}-\tilde\phi_\Lambda)\,,
\ee
where the angles $\tilde\phi$ are defined in the parton helicity frame~\cite{DAlesio:2021dcx}.

We can then compute the transverse $\Lambda$ polarization with respect to the jet-$\Lambda$ plane by setting
\be
\sin(\tilde\phi_{S_\Lambda}-\tilde\phi_\Lambda)=1\,.
\ee
This leads to the factorized cross section expressions,
\be
\label{eq:pol}
\begin{aligned}
d\Delta\sigma = \sum_{a,b,c,d}\int dx_adx_b\, \frac{\alpha_s^2}{\hat s}f_{a/A}(x_a) f_{b/B}(x_b)\\
\times\, |\overline{M}|^2  \delta(\hat s + \hat t +\hat u)\,\Delta D_{\Lambda^\uparrow/c}(\xi,p_{\perp\Lambda})
\end{aligned}
\ee
\be
\begin{aligned}
\label{eq:unpol}
d\sigma_{\rm unp} = \sum_{a,b,c,d}\int dx_adx_b\, \frac{\alpha_s^2}{\hat s}f_{a/A}(x_a) f_{b/B}(x_b)\\
\times\, |\overline{M}|^2  \delta(\hat s + \hat t +\hat u) \,D_{\Lambda/c}(\xi,p_{\perp\Lambda})\,,
\end{aligned}
\ee
where $D_{\Lambda/c}(\xi,p_{\perp\Lambda})$ is the unpolarized TMDFF.

By exploiting the delta-function in Eqs.~\eqref{eq:pol} and \eqref{eq:unpol}, one can integrate over one of the partonic variables $x_a$, $x_b$, so that they become
\be
\begin{aligned}
d\Delta\sigma = \sum_{a,b,c,d}\int \frac{dx_a}{x_a s - \sqrt s E_{\rm j}(1+\cos\theta_{\rm j})}\, \frac{\alpha_s^2}{\hat s}\\
\times\,f_{a/A}(x_a) f_{b/B}(x_b)\,
|\overline{M}|^2  \,\Delta D_{\Lambda^\uparrow/c}(\xi,p_{\perp\Lambda})\\
d\sigma_{\rm unp} = \sum_{a,b,c,d}\int \frac{dx_a}{x_a s - \sqrt s E_{\rm j}(1+\cos\theta_{\rm j})}\, \frac{\alpha_s^2}{\hat s}\\
\times\,f_{a/A}(x_a) f_{b/B}(x_b)\,|\overline{M}|^2  \,D_{\Lambda/c}(\xi,p_{\perp\Lambda})\,,
\end{aligned}
\ee
with
\be
x_b = \frac{x_a E_{\rm j} (1-\cos\theta_{\rm j})}{x_a \sqrt s - E_{\rm j}(1+\cos\theta_{\rm j})}\,.
\ee
Notice that in the above equations we will use the TMDFFs (both the unpolarized and the polarizing one) within the CSS framework,  presented in Refs.~\cite{DAlesio:2022brl, DAlesio:2023ozw}, by inverse Fourier transforming from $b_T$ to $k_T$ space, as used here.

For a massive hadron, one can define further several scaling variables:
\bea
z_\Lambda &=& E_\Lambda/E_{\rm j}\quad ({\rm energy \; fraction})\\
z_p &=& |\bm{p}_\Lambda|/E_{\rm j} \quad ({\rm momentum \; fraction})
\eea
or, as adopted in the experimental analysis we are going to consider,
\be
z = \frac{\bm{p}_\Lambda \cdot \bm{p}_{\rm j}}{\bm{p}_{\rm j}^2} =  \frac{\bm{p}_\Lambda \cdot \hat{\bm{p}}_{\rm j}}{E_{\rm j}} = \frac{\tilde p_{L\Lambda}}{E_{\rm j}}\,,
\ee
where $ \tilde p_{L\Lambda}$ is the longitudinal momentum of the $\Lambda$ along the jet direction.
This scaling variable can be directly related to the light-cone momentum fraction $\xi$ as follows:
\be
\begin{aligned}
\xi = \frac{E_\Lambda + \tilde p_{L\Lambda}}{2E_{\rm j}} =
& \, \frac{1}{2} \Big[\sqrt{z^2 +(p_{\perp\Lambda}^2 + m_{\Lambda}^2)/E_{\rm j}^2} +z \Big]\,.
\end{aligned}
\ee
We can then express the transverse $\Lambda$ polarization in Eq.~(\ref{polT}) as a function of $z$ (instead of $\xi$) by using
\bea
\frac{d\Delta\sigma(\bm{p}_{\rm j}, z,\bm{p}_{\perp\Lambda})}{dz} &=& \frac{d\xi}{dz} \,\frac{d\Delta\sigma(\bm{p}_{\rm j}, \xi,\bm{p}_{\perp\Lambda})}{d\xi} \\
\frac{d\sigma_{\rm unp}(\bm{p}_{\rm j}, z,\bm{p}_{\perp\Lambda})}{dz} &=& \frac{d\xi}{dz} \,\frac{d\sigma_{\rm unp}(\bm{p}_{\rm j}, \xi,\bm{p}_{\perp\Lambda})}{d\xi}\,,
\eea
with
\be
\frac{d\xi}{dz} = \frac{1}{2} \Big[ 1+ 1/\sqrt{1 +(p_{\perp\Lambda}^2 + m_{\Lambda}^2)/(z E_{\rm j})^2}
\Big]\,.
\ee
Notice that this extra factor simplifies in the polarization observable (ratio of cross sections) only at fixed kinematical variables, while it could play a role when one integrates the cross sections over $z$ and/or $p_{\perp\Lambda}$.

\section{Phenomenology}
\label{4_phenomenology}

Here we present estimates for the transverse $\Lambda/\bar\Lambda$ polarization in $pp\to {\rm jet}\, \Lambda^\uparrow\, X$ at the center-of-mass energy $\sqrt s=200$~GeV and compare them against STAR preliminary data~\cite{Gao:2024dxl}.
The kinematic cuts adopted in the experimental analysis are:
\begin{align}\label{expcuts}
&p_{\perp\Lambda} \le 1.6\,{\rm GeV}/c, \quad 0\le z\le 1, \nonumber\\
&8 \le p_{\rm{j}T} \le 25\, {\rm GeV}/c \quad
{\rm with}\quad
\langle p_{\rm{j}T} \rangle = 11 \, {\rm GeV}/c, \nonumber\\
 &|\eta_{\rm j}| \le 1.0, \,\,
 p_{T\Lambda} \le 10\, {\rm GeV}/c, \,\, |\eta_{\Lambda}| \le 1.5,
\end{align}
where $p_{\perp\Lambda}$ coincides with $j_T$, as adopted by the STAR Collaboration.

Our estimates will be computed at fixed $\eta_{\rm j}=0$ and $p_{\rm{j}T} = 11$ GeV/$c$. The latter, being the hard energy scale of the process, will be used as factorization scale. We will discuss this choice in more detail below.

Further, when we integrate over $z$ we limit to the region $z<0.8$, and when we integrate over $p_{\perp\Lambda}$, we limit to the region $p_{\perp\Lambda}\le 1.2$ GeV/$c$. This is indeed the region effectively covered by the data and Monte Carlo simulations (see Refs.~\cite{Gao:2024dxl,spinGao}).
Another important constraint comes from the reconstruction of the jet. Following the experimental analysis we will consider the anti-$k_T$ algorithm with a jet-cone radius $R=0.6$.
For further details on the jet reconstruction and the transverse momentum distribution of hadrons within a jet see also Refs.~\cite{Cacciari:2008gp, Kaufmann:2015hma, Kang:2017btw}.

Before presenting our results, we summarize
the information already extracted on the polarizing FF by fitting Belle data on transverse $\Lambda$ polarization in $e^+e^-$ annihilation processes at $Q=10.58$~GeV~\cite{Belle:2018ttu}. It is worth noticing that this is almost equal to the scale we will adopt in the present analysis.

Two data sets are available: one for the associated production of $\Lambda$'s together with a light hadron in an almost back-to-back configuration, and one for the inclusive $\Lambda$ production with the reconstruction of the thrust axis in the opposite hemisphere.
It is important to stress that while
for the first case a well defined TMD factorization approach has been formally developed, the second one presents some subtleties, and maybe related to modified TMD factorization schemes.
The latter has been indeed discussed in a series of papers~\cite{Makris:2020ltr, Boglione:2020cwn, Boglione:2021wov, Boglione:2022nzq}, showing that the absence of a hadron in the second (opposite) hemisphere prevents the symmetric absorption of the soft radiation factor, resulting in a TMDFF with a more complex nonperturbative structure.

For this reason in Ref.~\cite{DAlesio:2023ozw} we performed a fit, within the CSS framework, limiting to the associated production data set and focusing at the same time on the $SU(2)$ symmetry issue. On the other hand, in Ref.~\cite{DAlesio:2022brl}, in an exploratory combined fit including both associated production and inclusive data sets, in order to obtain a reasonable description we had to adopt two different nonperturbative models for the polFF.
In this respect this was, and has to be considered, only as a first attempt, also because of some critical aspects of the inclusive data set used in the fit.

In the present analysis we will consider four different parameterizations of the polFFs; three scenarios (Sc.s 1, 2 and 3) based on the associated production data fit~\cite{DAlesio:2023ozw}, and one from the combined fit within a double model (DM) for the nonperturbative part~\cite{DAlesio:2022brl}:
\begin{enumerate}
    \item Three different polFFs for $u,d,s$ quarks, and a single one for the sea antiquarks ($\bar{u}=\bar{d}=\bar{s}$), no charm contribution in the unpolarized cross section and no use of $SU(2)$ isospin symmetry (Sc.~1);
    \item Same as in Sc.~1 but with the inclusion of the charm contribution in the unpolarized cross section (Sc.~2);
    \item Inclusion of the charm contribution in the unpolarized cross section and use of $SU(2)$ isospin symmetry for the $u$, $d$ quark polFFs, adopting different pFFs for $s$ and $\bar{s}$ quarks (Sc.~3);
    \item Same as in Sc.~1, from the combined fit and adopting the model parametrization of Ref.~\cite{DAlesio:2023ozw} which better describe the inclusive data set (DM in the following).
\end{enumerate}

This will allow us to explore and test several important issues: $i)$ the universality of the polarizing FF; $ii)$ the role of the charm contribution and $SU(2)$ isospin symmetry; $iii)$ the nature of these effects in $pp$ collisions with respect to the corresponding ones observed in $e^+e^-$ annihilation processes for associated production or in the inclusive case.

We stress once again that the DM parametrization (from the combined fit) has been adopted here only for completeness.

Another important remark is that, in contrast with $e^+e^-$ and SIDIS processes, in $pp$ collisions gluons enter at the same perturbative order as quarks. In other words, parton $c$ can be also a gluon and the two contributions in the fragmentation process add up together. This means that a nonzero gluon polarizing fragmentation function, still totally unknown, could contribute to the transverse $\Lambda$ polarization.
In the following, while keeping the gluon contribution in the denominator, via a nonzero unpolarized gluon TMDFF, we set the gluon polFF to zero. We will come back to this very interesting point below.

We must emphasize here that the unpolarized gluon TMDFF has not been extracted so far.  For it we will employ the same TMD structure as for quarks, while for the perturbative and nonperturbative Sudakov soft factors we adopt suitable expressions, as given in~\ref{Apx_A}.

Lastly, for what concerns the collinear parton distribution functions we will employ the next-to-next-to-leading order CT14 set~\cite{Dulat:2015mca} at the scale $\mu=p_{{\rm j}T}$.

\begin{center}
\begin{figure}[!t]
\includegraphics[width=7cm]
{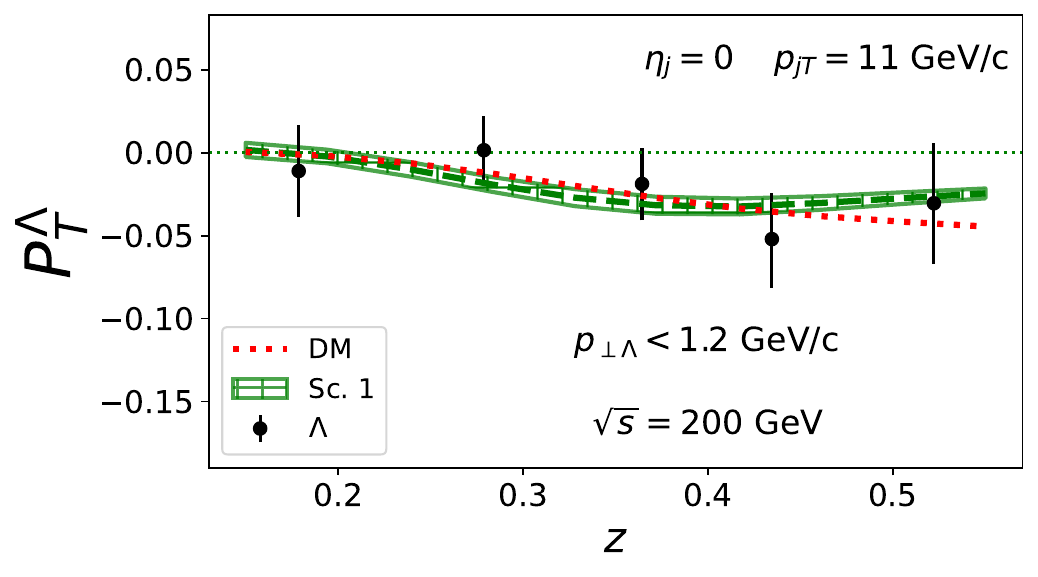}\hspace{.2cm}
\includegraphics[width=7cm]{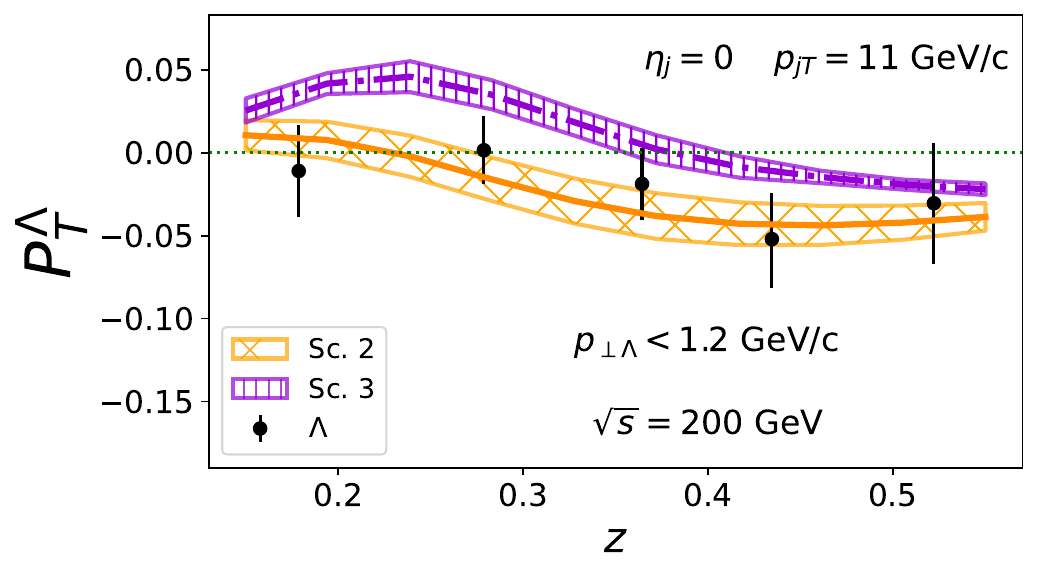}
\caption{Estimates of the transverse $\Lambda$ polarization in $pp\to {\rm jet}\, \Lambda\, X$ as a function of $z$ at $\sqrt s=200$~GeV, $\eta_{\rm j}=0$  and $p_{{\rm j}T}= 11$~GeV/c, adopting for the polFFs the parametrizations of Sc.~1 and DM  (upper panel), and those of Sc.s~2 and 3 (lower panel), see text.
Uncertainty bands at 2-$\sigma$ CL are also shown for Sc.s~1-3.
Preliminary STAR data are from Ref.~\cite{Gao:2024dxl}.}
\label{fig:Lambdaz}
\end{figure}
\end{center}

A more detailed comment on the choice of the factorization scale both in the collinear PDFs and the TMDFFs is mandatory.
In fact, the relevant scale for the TMDFFs for this kind of process is $\mu_j=p_{{\rm j}T} R$, where $R$ is the jet-cone radius, as discussed in Ref.~\cite{Kang:2017glf}. Then, by properly evolving up to $\mu=p_{{\rm j}T}$ one can resum single logarithms in the jet size parameter to all orders in $\alpha_s$, the strong coupling constant. Since our study is performed at LO accuracy, which is independent of $R$ (and more generally of the jet dynamics), we will use $\mu=p_{{\rm j}T}$ for the unpolarized and the polarizing TMDFFs and similarly for the collinear PDFs: the same procedure was implemented in Ref.~\cite{Kang:2017btw} where the authors studied the Collins asymmetry of hadrons in a jet.

\begin{center}
\begin{figure}[!t]
\includegraphics[width=7cm]{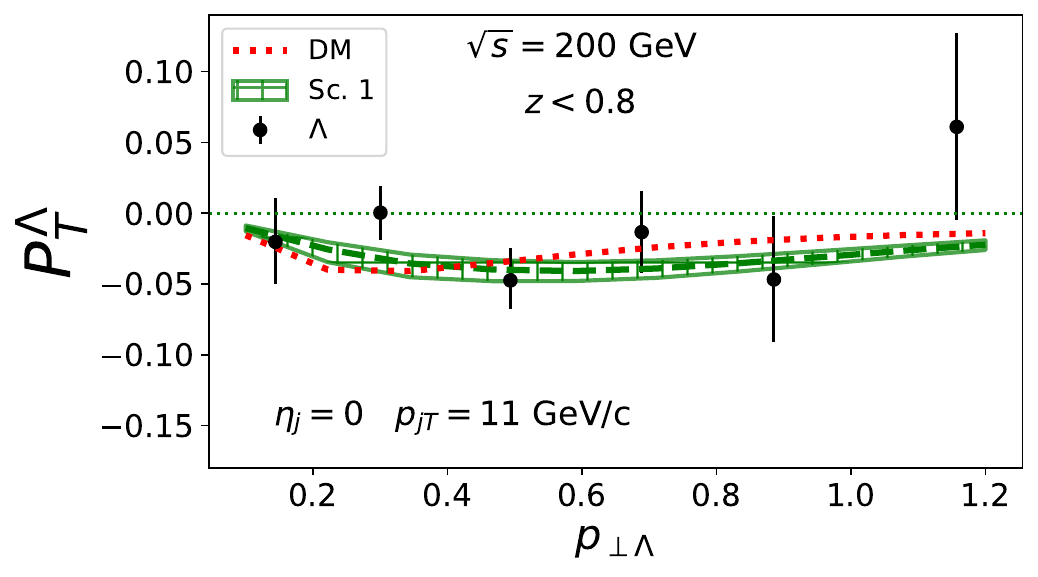}\hspace{.2cm}
\includegraphics[width=7cm]{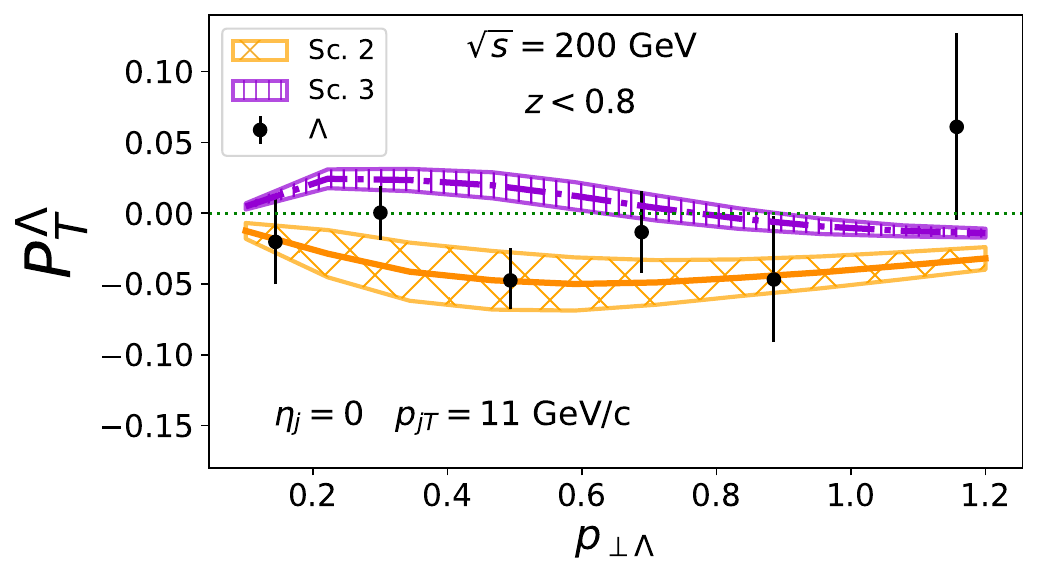}
\caption{Estimates of the transverse $\Lambda$ polarization in $pp\to {\rm jet}\, \Lambda\, X$ as a function of $p_{\perp\Lambda}$ at $\sqrt s=200$~GeV, $\eta_{\rm j}=0$  and $p_{{\rm j}T}= 11$~GeV/c, adopting for the polFFs the parametrizations of Sc.~1 and DM (upper panel), and those of Sc.s~2 and 3 (lower panel), see text. Uncertainty bands at 2-$\sigma$ CL are also shown for Sc.s~1-3. Preliminary STAR data are from Ref.~\cite{Gao:2024dxl}.}
\label{fig:LambdajT}
\end{figure}
\end{center}

\begin{center}
\begin{figure}[!t]
\includegraphics[width=7cm]
{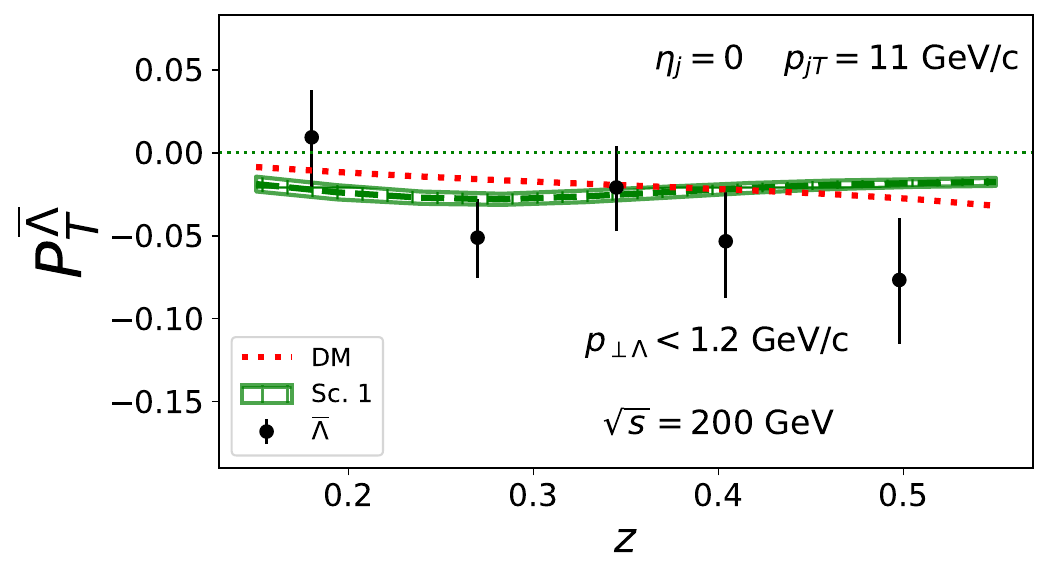}\hspace{.2cm}
\includegraphics[width=7cm]{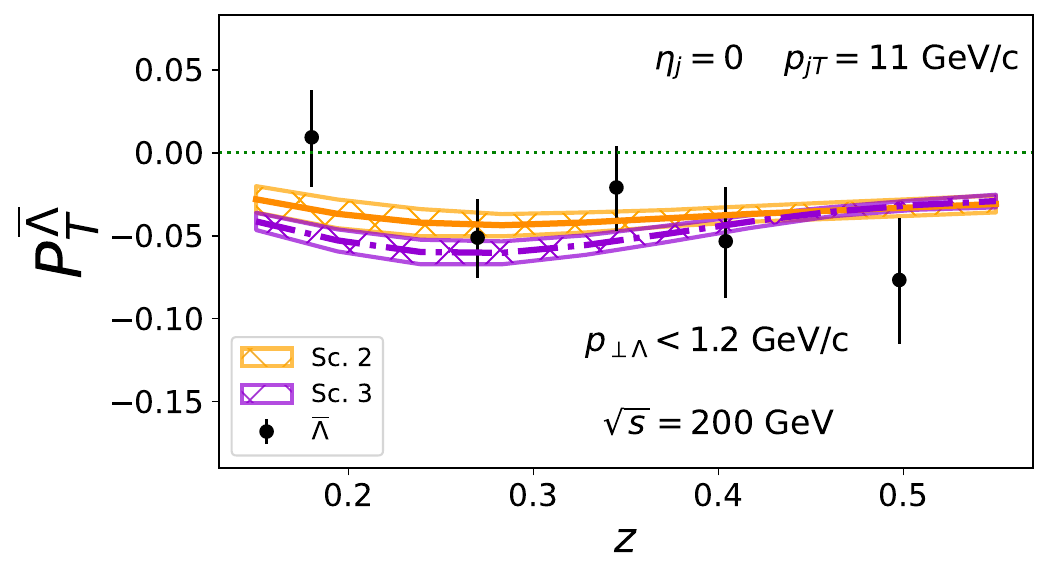}
\caption{Same as in Fig.~\ref{fig:Lambdaz} but for the transverse $\bar\Lambda$ polarization. }
\label{fig:Lambdabarz}
\end{figure}
\end{center}

\begin{center}
\begin{figure}[!t]
\includegraphics[width=7cm]{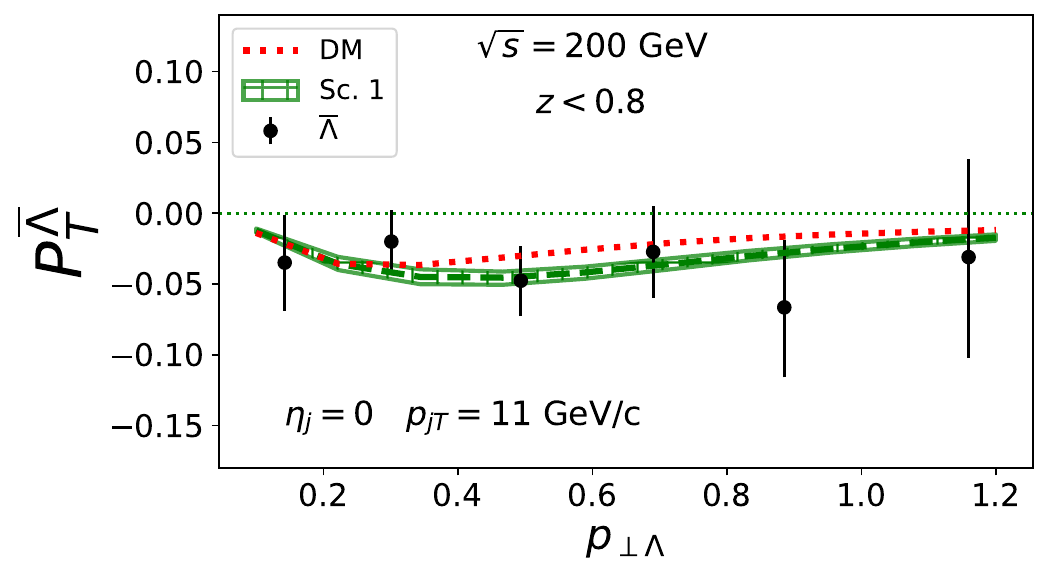}\hspace{.2cm}
\includegraphics[width=7cm]{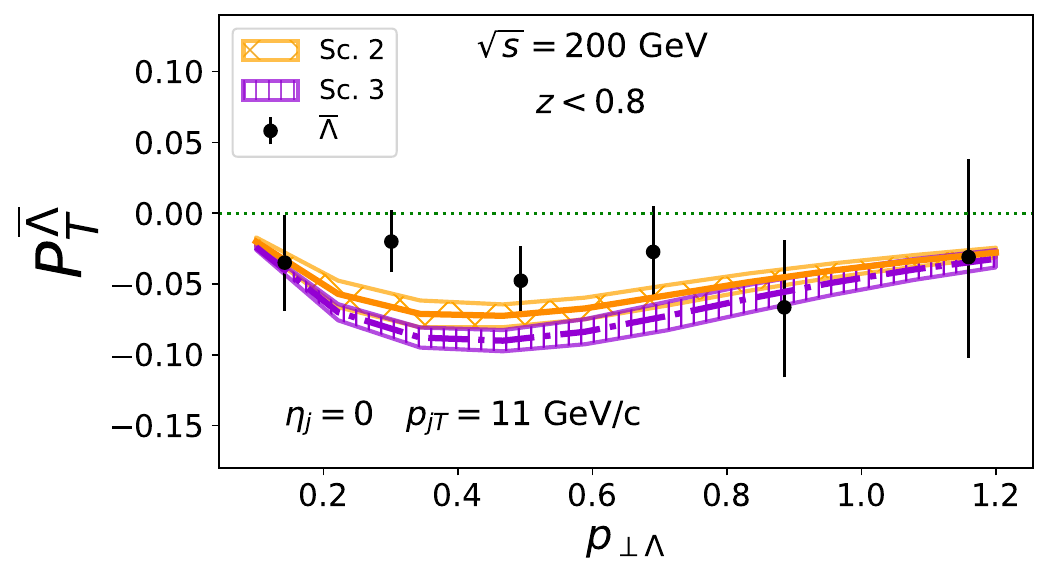}
\caption{Same as in Fig.~\ref{fig:LambdajT} but for the transverse $\bar\Lambda$ polarization.}
\label{fig:LambdabarjT}
\end{figure}
\end{center}

In Fig.~\ref{fig:Lambdaz} we show our predictions for the transverse $\Lambda$ polarization as a function of $z$ adopting Sc.~1 (green dashed lines) and DM (red dotted-line) parametrizations (upper panel), and Sc.s 2 (orange solid lines) and 3 (purple dot-dashed lines) (lower panel) against STAR preliminary data, while in Fig.~\ref{fig:LambdajT} we show the corresponding curves as a function of $p_{\perp\Lambda}$. In both cases we integrate over the other variable. For the three scenarios from the associated production fit we also show the statistical uncertainty bands at 2-$\sigma$ confidence level (CL) from the uncertainties in the polFFs as determined in Ref.~\cite{DAlesio:2023ozw}. We note that to speed up  the numerical computation, which involves inverse Fourier transforms, we employ a compression procedure to reduce the large number of polFF sets used to generate the bands (see Ref.~\cite{Boglione:2024dal}). Corresponding estimates for the transverse $\bar\Lambda$ polarization as a function of $z$ and $p_{\perp\Lambda}$ are shown respectively in Figs.~\ref{fig:Lambdabarz} and \ref{fig:LambdabarjT}.

A few comments are in order: $i)$ the behaviour in $z$ is driven by the relative contributions of the polFFs. More precisely, while within Sc.s 1 and 2 (and similarly for the DM case) only the up polFF is positive (see Fig.~4 of Ref.~\cite{DAlesio:2023ozw}), in Sc.~3 also the down polFF is positive (see Fig.~5 of Ref.~\cite{DAlesio:2023ozw}). In all cases they are strongly suppressed at large $z$. This turns into a positive value of the polarization for $\Lambda$ in Sc.~3 at small $z$, becoming negative at intermediate/large $z$, and negative values for Sc.s~1, 2 and DM over almost the entire $z$ range. For $\bar\Lambda$ the negative values within all scenarios are driven by the negative sign of the sea polFFs, coupled with the up and down valence content of the incoming protons. Notice that the minimum value of the initial parton light-cone momentum fractions, $x_a,x_b$, explored in this kinematical configuration is around 0.06. Similar considerations are valid for the $p_{\perp\Lambda}$ behaviour (see Figs.~\ref{fig:LambdajT} and \ref{fig:LambdabarjT}).

For completeness we have also considered the role of intrinsic charm in the proton, an issue already discussed in Ref.~\cite{DAlesio:2023ozw} in the context of the corresponding estimates for SIDIS processes. In such a case the results for the transverse polarization are almost indistinguishable from the curves shown above.

As one can see from the plots, the large experimental error bars prevent us to draw strong conclusions and/or to adopt or disregard any of the scenarios considered. We can nevertheless observe a general agreement with data, in favour of the predicted universality property of the polFF.
More precisely, concerning the three scenarios obtained from the associated production fits, only one scenario, Sc.~3, gives estimates (purple solid lines/bands) that are somewhat far from the data.
On the other hand, Sc.~1 (green dashed lines/bands) and, to a lesser extent, Sc.~2 (orange dot-dashed lines/bands) seem to be able to describe the data fairly well. It is important to note that Sc.~1 has been extracted without considering the non-negligible contribution from the charm quark fragmentation into $\Lambda$'s in $e^+e^-$ processes. In contrast, in Sc.~2 (as well as in Sc.~3), for which this contribution has been included in the unpolarized cross section (the denominator of the transverse polarization), the corresponding term driven by the polFF for charm quark (in the numerator) was not taken into account (see Ref.~\cite{DAlesio:2023ozw} for more details).
In this respect, this open issue has to be properly addressed in the future.

Moreover, as we will discuss below, there is another totally unknown, and potentially non-negligible, contribution coming from the gluon polFF. Finally, for what concerns the estimates from the double-model fit (red dotted lines), which shares the same flavor structure as for Sc.~1, there are, as already stated, even more fundamental issues still to be addressed.

We  mention that an analogous phenomenological study was carried out in Ref.~\cite{Kang:2020xyq}, where predictions for RHIC kinematics around the central rapidity region were given. The authors employed the TMDJFFs, properly connected to the TMD polFFs (corresponding to our Sc.~1), as extracted from associated production $e^+e^-$ data.
It is noteworthy that these analyses come to similar results, indicating that the LO framework is well motivated.

As a final remark we come back to the potential role of the gluon polFF. We have checked that in this kinematical region the contribution to the unpolarized $\Lambda$ and $\bar\Lambda$ cross section coming from gluon fragmentation is about 50\%, both as a function of $z$ and $p_{\perp\Lambda}$.
This implies that, since the estimates from the quark contribution to the polarization are around 5-8\% in size (at most) and almost compatible with data in all scenarios, only a gluon polFF reduced in size at about 10\% of its positivity bound  would be allowed.
Even if only on a qualitative level, this is the first hint ever on the size of the gluon polFF based on available data. We stress once again that this is possible because in this process, at variance with  $e^+e^-$ and SIDIS processes, gluons and quarks enter at the same perturbative order.
Last but not least, even such a reduced contribution could help in improving the agreement with data.
At this stage, the large error bars prevent one to further exploit this issue. Future and improved experimental analyses will be extremely helpful in this respect.

\section{Conclusions}
\label{5_conclusions}

In this paper we have presented a phenomenological analysis of the transverse polarization of $\Lambda$ and $\bar\Lambda$ hyperons within a jet, produced in unpolarized proton-proton collisions. Adopting a hybrid approach, with a collinear factorization scheme in the initial state and keeping TMD effects only in the fragmentation mechanism, we have presented several theoretical estimates, based on recent extractions of the quark polarizing fragmentation functions from fits to Belle $e^+e^-$ data for the associated and inclusive $\Lambda$ production.
By comparing these predictions with recent preliminary data by the STAR Collaboration in $pp$ collisions, we have carried out the first attempt to test the universality of the polFF.

Although the large error bars still prevent one to draw any definite conclusions, the present estimates are compatible with the measured data points, corroborating the expected universality property of T-odd TMDFFs.

As a by-product we have elaborated on the totally unknown gluon polFF, that in principle enters at the same perturbative order as the quarks. Based on the present data, even if within their large uncertainties, some hints towards a strong reduction in its size, by around 10\% with respect to the positivity bound, has been inferred. This process could then represent the golden channel to get information on this totally unknown TMDFF.

New and more precise data, maybe also at different energies to access  larger hard scales, will definitely be  useful in refining this analysis and testing, together with the universality, the scale evolution of the polFFs. Moreover,
they will surely play a significant role in disentangling the nature of spontaneous transverse polarization of $\Lambda$ hyperons and its connection to what has been observed in $e^+e^-$ processes and to future SIDIS measurements at the Electron Ion Collider~\cite{AbdulKhalek:2021gbh,Kang:2021ffh,Kang:2021kpt,DAlesio:2023ozw}.

\section*{Acknowledgments}
We thank Taoya Gao and Qinghua Xu for their help on some aspects of the STAR experimental analysis. We acknowledge Carlo Flore for his support in the computation of the uncertainty bands. This project has received funding from the European Union’s Horizon 2020 research and innovation programme under grant agreement N.~824093 (STRONG-2020).  U.D.~and M.Z.~also acknowledge financial support by Fondazione di Sardegna under the project ``Matter-antimatter asymmetry and polarisation in strange hadrons at LHCb'', project number F73C22001150007 (University of Cagliari).
L.G.~acknowledges support from the US Department of Energy under contract No.~DE-FG02-07ER41460 and from the INFN Cagliari division, and the kind hospitality offered to him by the Physics Department of the University of Cagliari during the development of this work. This project was partially supported by the U.S.~Department of Energy, Office of Science, Contract No.~DE-AC05-06OR23177, under which Jefferson Science Associates, LLC operates Jefferson Lab.

\appendix

\section{TMDFFs and Sudakov soft factors}
\label{Apx_A}

The full expression in $b_T$ space for the unpolarized TMDFF for quarks and gluons, as adopted in Ref.~\cite{DAlesio:2022brl} (to which we refer for all details), is given by
\begin{multline}
\widetilde{D}_{1, h/i}(\xi,b_c(b_T);Q^2,Q) =
\frac{1}{\xi^2}
d_{h/i}(\xi; \bar{\mu}_b)\, M_D(b_c(b_T),\xi) \\
\times  \exp\bigg\{  - g_K^i (b_c(b_T);b_{\text{max}})\ln\frac{Q \xi}{M_{h}} +\frac{1}{2}S_{\rm pert}^i(b_*;\bar{\mu}_b)  \bigg\} \,,
\label{D_full_bc_Q}
\end{multline}
with $i=q,g$, where we have used the LO expression for the coefficient function~\cite{DAlesio:2022brl} and where $S_{\rm pert}^i$ is the perturbative Sudakov soft factor. Its analytic expression is given as
\begin{equation}
\begin{aligned}
    S^i_{\rm pert}(b_*;\bar{\mu}_{b})  &= \widetilde{K}^i(b_*;\bar{\mu}_{b}) \ln\frac{Q^2}{\bar{\mu}_{b}^2} \\
    &\hspace*{-0.2cm}+\int^{Q}_{\bar{\mu}_{b}} \frac{d\mu'}{\mu'}\,\bigg[ 2\gamma_D^i(g(\mu');1) - \gamma_K^i(g(\mu')) \ln{\frac{Q^2}{\mu'^2}} \bigg] \,.
\end{aligned}
\end{equation}
At next-to-leading-logarithmic accuracy
we take $\alpha_s$ at LO order:
\begin{equation}
        \alpha_s(\mu^2) = \frac{1}{\beta_0 \ln(\mu^2/\Lambda^2_{\rm QCD})}\,,
\end{equation}
and we expand the cusp and non-cusp anomalous dimensions as follows:
\begin{equation}
    \gamma_K^i = \sum_n  \gamma^{i[n]}_K \bigg(\frac{\alpha_s}{4\pi}\bigg)^n \, \quad \gamma_{D}^i = \sum_n  \gamma^{i[n]}_{D} \bigg(\frac{\alpha_s}{4\pi}\bigg)^n \,,
\end{equation}
retaining up to, respectively, the second and first order.
Given that the first order term of $\widetilde{K}(b_*;\bar{\mu}_{b})$ is zero~\cite{collins_2011,Aybat:2011zv}, the perturbative Sudakov factor can be written again as:
\begin{multline}
 S^i_{\rm pert}(b_*;\bar{\mu}_{b})=\frac{\gamma^{i[1]}_D}{4\pi \beta_0}\ln\bigg(\frac{\ln(Q/\Lambda_{\rm QCD})}{\ln(\bar{\mu}_{b}/\Lambda_{\rm QCD})}\bigg) \\
    +\frac{\gamma^{i[1]}_K}{4\pi \beta_0}\bigg[\ln(Q/\bar{\mu}_{b})- \ln(Q/\Lambda_{\rm QCD}) \ln\bigg(\frac{\ln(Q/\Lambda_{\rm QCD})}{\ln(\bar{\mu}_{b}/\Lambda_{\rm QCD})}\bigg) \bigg]\nonumber \\
    + \frac{\gamma^{i[2]}_K}{2(4\pi \beta_0)^2} \bigg[- \frac{\ln(Q/\bar{\mu}_{b})}{\ln(\bar{\mu}_{b}/\Lambda_{\rm QCD})} + \ln\bigg(\frac{\ln(Q/\Lambda_{\rm QCD})}{\ln(\bar{\mu}_{b}/\Lambda_{\rm QCD})}\bigg)\bigg] \,,
\label{eq:spert_def_res}
\end{multline}
where~\cite{Collins:2017oxh, Echevarria:2016scs}:
\begin{multline}
\beta_0 = \frac{11 C_A - 4 T_F n_f}{12 \pi}\\
\gamma^{q[1]}_{D}=6 C_F  \quad \gamma^{g[1]}_{D}=\frac{22}{3}C_A -\frac{8}{3} T_F n_f \\
    \gamma^{q[1]}_K = 8 C_F \,, \; \gamma^{q[2]}_K = C_A C_F \bigg(\frac{536}{9} -  \frac{8  \pi^2}{3}\bigg) - \frac{80}{9} C_F n_f \\
    \gamma^{g[1]}_K = 8 C_A \,, \; \gamma^{g[2]}_K = C_A^2
    \bigg(\frac{536}{9} -  \frac{8  \pi^2}{3}\bigg) - \frac{80}{9} C_A n_f
    \,,
\end{multline}
with $C_F=4/3$, $C_A=3$, $T_F=1/2$, and $\Lambda_{\rm QCD}=0.2123 \,\text{GeV}$ for $n_f=3$ or $\Lambda_{\rm QCD}=0.1737 \,\text{GeV}$ for $n_f=4$.

For the nonperturbative Sudakov soft factor we adopt the Pavia extraction for the quark TMDFF~\cite{Bacchetta:2017gcc}, while for the gluon TMDFF we rescale it by a factor $C_A/C_F$, as discussed in Ref.~\cite{Kang:2017glf}
\bea
g_K^q &= & g_2\frac{b_T^2}{2}  \qquad  g_2=0.13 \;{\rm GeV}^2\\
g_K^g &=& \frac{C_A}{C_F} g_K^q\,,
\eea
employing the same quark nonperturbative part, $M_D$.


\begin{thebibliography}{10}
\expandafter\ifx\csname url\endcsname\relax
  \def\url#1{\texttt{#1}}\fi
\expandafter\ifx\csname urlprefix\endcsname\relax\def\urlprefix{URL }\fi
\expandafter\ifx\csname href\endcsname\relax
  \def\href#1#2{#2} \def\path#1{#1}\fi

\bibitem{Collins:1981uk}
J.~C. Collins, D.~E. Soper, {Back-to-back jets in QCD}, Nucl. Phys. B 193
  (1981) 381, [Erratum: Nucl. Phys. B 213 (1983) 545].
\newblock \href {https://doi.org/https://doi.org/10.1016/0550-3213(81)90339-4}
  {\path{doi:https://doi.org/10.1016/0550-3213(81)90339-4}}.

\bibitem{Collins:1981va}
J.~C. Collins, D.~E. Soper, {Back-to-back jets: Fourier transform from $b$ to
  $k_T$}, Nucl. Phys. B 197 (1982) 446.
\newblock \href {https://doi.org/https://doi.org/10.1016/0550-3213(82)90453-9}
  {\path{doi:https://doi.org/10.1016/0550-3213(82)90453-9}}.

\bibitem{Collins:1984kg}
J.~C. Collins, D.~E. Soper, G.~Sterman, {Transverse momentum distribution in
  Drell-Yan pair and $W$ and $Z$ boson production}, Nucl. Phys. B 250 (1985)
  199.
\newblock \href {https://doi.org/10.1016/0550-3213(85)90479-1}
  {\path{doi:10.1016/0550-3213(85)90479-1}}.

\bibitem{collins_2011}
J.~Collins, Foundations of Perturbative QCD, Cambridge Monographs on Particle
  Physics, Nuclear Physics and Cosmology, Cambridge University Press, 2011.
\newblock \href {https://doi.org/10.1017/CBO9780511975592}
  {\path{doi:10.1017/CBO9780511975592}}.

\bibitem{Boussarie:2023izj}
R.~Boussarie, et~al., {TMD Handbook} (2023).
\newblock \href {http://arxiv.org/abs/2304.03302} {\path{arXiv:2304.03302}}.

\bibitem{Metz:2002iz}
A.~Metz, {Gluon-exchange in spin-dependent fragmentation}, Phys. Lett. B 549
  (2002) 139--145.
\newblock \href {http://arxiv.org/abs/hep-ph/0209054}
  {\path{arXiv:hep-ph/0209054}}, \href
  {https://doi.org/10.1016/S0370-2693(02)02899-X}
  {\path{doi:10.1016/S0370-2693(02)02899-X}}.

\bibitem{Boer:2003cm}
D.~Boer, P.~J. Mulders, F.~Pijlman, {Universality of T-odd effects in single
  spin and azimuthal asymmetries}, Nucl. Phys. B 667 (2003) 201.
\newblock \href {http://arxiv.org/abs/hep-ph/0303034}
  {\path{arXiv:hep-ph/0303034}}, \href
  {https://doi.org/10.1016/S0550-3213(03)00527-3}
  {\path{doi:10.1016/S0550-3213(03)00527-3}}.

\bibitem{Collins:2004nx}
J.~C. Collins, A.~Metz, {Universality of soft and collinear factors in
  hard-scattering factorization}, Phys. Rev. Lett. 93 (2004) 252001.
\newblock \href {http://arxiv.org/abs/hep-ph/0408249}
  {\path{arXiv:hep-ph/0408249}}, \href
  {https://doi.org/10.1103/PhysRevLett.93.252001}
  {\path{doi:10.1103/PhysRevLett.93.252001}}.

\bibitem{Yuan:2009dw}
F.~Yuan, J.~Zhou, {Collins Fragmentation and the Single Transverse Spin
  Asymmetry}, Phys. Rev. Lett. 103 (2009) 052001.
\newblock \href {http://arxiv.org/abs/0903.4680} {\path{arXiv:0903.4680}},
  \href {https://doi.org/10.1103/PhysRevLett.103.052001}
  {\path{doi:10.1103/PhysRevLett.103.052001}}.

\bibitem{Boer:2010ya}
D.~Boer, Z.-B. Kang, W.~Vogelsang, F.~Yuan, {Test of the Universality of
  Naive-time-reversal-odd Fragmentation Functions}, Phys. Rev. Lett. 105 (2010)
  202001.
\newblock \href {http://arxiv.org/abs/1008.3543} {\path{arXiv:1008.3543}},
  \href {https://doi.org/10.1103/PhysRevLett.105.202001}
  {\path{doi:10.1103/PhysRevLett.105.202001}}.

\bibitem{Collins:1992kk}
J.~C. Collins, {Fragmentation of transversely polarized quarks probed in
  transverse momentum distributions}, Nucl. Phys. B 396 (1993) 161.
\newblock \href {http://arxiv.org/abs/hep-ph/9208213}
  {\path{arXiv:hep-ph/9208213}}, \href
  {https://doi.org/10.1016/0550-3213(93)90262-N}
  {\path{doi:10.1016/0550-3213(93)90262-N}}.

\bibitem{Mulders:1995dh}
P.~J. Mulders, R.~D. Tangerman, {The complete tree level result up to order
  $1/Q$ for polarized deep-inelastic leptoproduction}, Nucl. Phys. B 461 (1996)
  197, [Erratum: Nucl. Phys. B 484, (1997) 538].
\newblock \href {http://arxiv.org/abs/hep-ph/9510301}
  {\path{arXiv:hep-ph/9510301}}, \href
  {https://doi.org/10.1016/0550-3213(95)00632-X}
  {\path{doi:10.1016/0550-3213(95)00632-X}}.

\bibitem{Anselmino:2000vs}
M.~Anselmino, D.~Boer, U.~D'Alesio, F.~Murgia, {$\Lambda$ polarization from
  unpolarized quark fragmentation}, Phys. Rev. D 63 (2001) 054029.
\newblock \href {http://arxiv.org/abs/hep-ph/0008186}
  {\path{arXiv:hep-ph/0008186}}, \href
  {https://doi.org/10.1103/PhysRevD.63.054029}
  {\path{doi:10.1103/PhysRevD.63.054029}}.

\bibitem{Yuan:2007nd}
F.~Yuan, {Azimuthal Asymmetric Distribution of Hadrons Inside a Jet at Hadron
  Collider}, Phys. Rev. Lett. 100 (2008) 032003.
\newblock \href {http://arxiv.org/abs/0709.3272} {\path{arXiv:0709.3272}},
  \href {https://doi.org/10.1103/PhysRevLett.100.032003}
  {\path{doi:10.1103/PhysRevLett.100.032003}}.

\bibitem{Meissner:2008yf}
S.~Meissner, A.~Metz, {Partonic pole matrix elements for fragmentation}, Phys.
  Rev. Lett. 102 (2009) 172003.
\newblock \href {http://arxiv.org/abs/0812.3783} {\path{arXiv:0812.3783}},
  \href {https://doi.org/10.1103/PhysRevLett.102.172003}
  {\path{doi:10.1103/PhysRevLett.102.172003}}.

\bibitem{Gamberg:2008yt}
L.~P. Gamberg, A.~Mukherjee, P.~J. Mulders, {Spectral analysis of gluonic pole
  matrix elements for fragmentation}, Phys. Rev. D 77 (2008) 114026.
\newblock \href {http://arxiv.org/abs/0803.2632} {\path{arXiv:0803.2632}},
  \href {https://doi.org/10.1103/PhysRevD.77.114026}
  {\path{doi:10.1103/PhysRevD.77.114026}}.

\bibitem{Gamberg:2010uw}
L.~P. Gamberg, A.~Mukherjee, P.~J. Mulders, {A model independent analysis of
  gluonic pole matrix elements and universality of TMD fragmentation
  functions}, Phys. Rev. D 83 (2011) 071503.
\newblock \href {http://arxiv.org/abs/1010.4556} {\path{arXiv:1010.4556}},
  \href {https://doi.org/10.1103/PhysRevD.83.071503}
  {\path{doi:10.1103/PhysRevD.83.071503}}.

\bibitem{Sivers:1989cc}
D.~W. Sivers, {Single Spin Production Asymmetries from the Hard Scattering of
  Point-Like Constituents}, Phys. Rev. D 41 (1990) 83.
\newblock \href {https://doi.org/10.1103/PhysRevD.41.83}
  {\path{doi:10.1103/PhysRevD.41.83}}.

\bibitem{Boer:1999mm}
D.~Boer, {Investigating the origins of transverse spin asymmetries at RHIC},
  Phys. Rev. D 60 (1999) 014012.
\newblock \href {http://arxiv.org/abs/hep-ph/9902255}
  {\path{arXiv:hep-ph/9902255}}, \href
  {https://doi.org/10.1103/PhysRevD.60.014012}
  {\path{doi:10.1103/PhysRevD.60.014012}}.

\bibitem{Collins:2002kn}
J.~C. Collins, {Leading twist single transverse-spin asymmetries: Drell-Yan and
  deep inelastic scattering}, Phys. Lett. B 536 (2002) 43--48.
\newblock \href {http://arxiv.org/abs/hep-ph/0204004}
  {\path{arXiv:hep-ph/0204004}}, \href
  {https://doi.org/10.1016/S0370-2693(02)01819-1}
  {\path{doi:10.1016/S0370-2693(02)01819-1}}.

\bibitem{Brodsky:2002rv}
S.~J. Brodsky, D.~S. Hwang, I.~Schmidt, {Initial-state interactions and
  single-spin asymmetries in Drell-Yan processes}, Nucl. Phys. B642 (2002)
  344--356.
\newblock \href {http://arxiv.org/abs/hep-ph/0206259}
  {\path{arXiv:hep-ph/0206259}}, \href
  {https://doi.org/10.1016/S0550-3213(02)00617-X}
  {\path{doi:10.1016/S0550-3213(02)00617-X}}.

\bibitem{Anselmino:2001js}
M.~Anselmino, D.~Boer, U.~D'Alesio, F.~Murgia, {Transverse $\Lambda$
  polarization in semiinclusive DIS}, Phys. Rev. D 65 (2002) 114014.
\newblock \href {http://arxiv.org/abs/hep-ph/0109186}
  {\path{arXiv:hep-ph/0109186}}, \href
  {https://doi.org/10.1103/PhysRevD.65.114014}
  {\path{doi:10.1103/PhysRevD.65.114014}}.

\bibitem{Belle:2018ttu}
Y.~Guan, et~al., Belle Collaboration, {Observation of Transverse $\Lambda/\bar{\Lambda}$ Hyperon
  Polarization in $e^+e^-$ Annihilation at Belle}, Phys. Rev. Lett. 122
  (2019) 042001.
\newblock \href {http://arxiv.org/abs/1808.05000} {\path{arXiv:1808.05000}},
  \href {https://doi.org/10.1103/PhysRevLett.122.042001}
  {\path{doi:10.1103/PhysRevLett.122.042001}}.

\bibitem{DAlesio:2020wjq}
U.~D'Alesio, F.~Murgia, M.~Zaccheddu, {First extraction of the $\Lambda$
  polarizing fragmentation function from Belle $e^+e^-$ data}, Phys. Rev. D
  102 (2020) 054001.
\newblock \href {http://arxiv.org/abs/2003.01128} {\path{arXiv:2003.01128}},
  \href {https://doi.org/10.1103/PhysRevD.102.054001}
  {\path{doi:10.1103/PhysRevD.102.054001}}.

\bibitem{Callos:2020qtu}
D.~Callos, Z.-B. Kang, J.~Terry, {Extracting the transverse momentum dependent
  polarizing fragmentation functions}, Phys. Rev. D 102 (2020) 096007.
\newblock \href {http://arxiv.org/abs/2003.04828} {\path{arXiv:2003.04828}},
  \href {https://doi.org/10.1103/PhysRevD.102.096007}
  {\path{doi:10.1103/PhysRevD.102.096007}}.

\bibitem{Gamberg:2021iat}
L.~Gamberg, Z.-B. Kang, D.~Y. Shao, J.~Terry, F.~Zhao, {Transverse $\Lambda$
  polarization in $e^+e^-$ collisions}, Phys. Lett. B 818 (2021) 136371.
\newblock \href {http://arxiv.org/abs/2102.05553} {\path{arXiv:2102.05553}},
  \href {https://doi.org/10.1016/j.physletb.2021.136371}
  {\path{doi:10.1016/j.physletb.2021.136371}}.

\bibitem{Kang:2021kpt}
Z.-B. Kang, J.~Terry, A.~Vossen, Q.~Xu, J.~Zhang, {Transverse Lambda production
  at the future Electron-Ion Collider}, Phys. Rev. D 105 (2022) 094033.
\newblock \href {http://arxiv.org/abs/2108.05383} {\path{arXiv:2108.05383}},
  \href {https://doi.org/10.1103/PhysRevD.105.094033}
  {\path{doi:10.1103/PhysRevD.105.094033}}.

\bibitem{Li:2020oto}
H.~Li, X.~Wang, Y.~Yang, Z.~Lu, {The transverse polarization of $\Lambda$
  hyperons in $e^+e^-\rightarrow \Lambda ^\uparrow h X$ processes within TMD
  factorization}, Eur. Phys. J. C 81 (2021) 289.
\newblock \href {http://arxiv.org/abs/2009.07193} {\path{arXiv:2009.07193}},
  \href {https://doi.org/10.1140/epjc/s10052-021-09064-1}
  {\path{doi:10.1140/epjc/s10052-021-09064-1}}.

\bibitem{Chen:2021hdn}
K.-B. Chen, Z.-T. Liang, Y.-L. Pan, Y.-K. Song, S.-Y. Wei, {Isospin symmetry of
  fragmentation functions}, Phys. Lett. B 816 (2021) 136217.
\newblock \href {http://arxiv.org/abs/2102.00658} {\path{arXiv:2102.00658}},
  \href {https://doi.org/10.1016/j.physletb.2021.136217}
  {\path{doi:10.1016/j.physletb.2021.136217}}.

\bibitem{DAlesio:2022brl}
U.~D'Alesio, L.~Gamberg, F.~Murgia, M.~Zaccheddu, {Transverse $\Lambda$
  polarization in $e^+e^-$ processes within a TMD factorization approach and
  the polarizing fragmentation function}, JHEP 12 (2022) 074.
\newblock \href {http://arxiv.org/abs/2209.11670} {\path{arXiv:2209.11670}},
  \href {https://doi.org/10.1007/JHEP12(2022)074}
  {\path{doi:10.1007/JHEP12(2022)074}}.

\bibitem{DAlesio:2023ozw}
U.~D'Alesio, L.~Gamberg, F.~Murgia, M.~Zaccheddu, {Transverse
  \ensuremath{\Lambda} polarization in $e^+e^-$ annihilations and in SIDIS
  processes at the EIC within TMD factorization}, Phys. Rev. D 108 (2023)
  094004.
\newblock \href {http://arxiv.org/abs/2307.02359} {\path{arXiv:2307.02359}},
  \href {https://doi.org/10.1103/PhysRevD.108.094004}
  {\path{doi:10.1103/PhysRevD.108.094004}}.

\bibitem{Chen:2021zrr}
K.-b. Chen, Z.-T. Liang, Y.-K. Song, S.-Y. Wei, {Longitudinal and transverse
  polarizations of $\Lambda$ hyperon in unpolarized SIDIS and $e^+e^-$
  annihilation}, Phys. Rev. D 105 (2022) 034027.
\newblock \href {http://arxiv.org/abs/2108.07740} {\path{arXiv:2108.07740}},
  \href {https://doi.org/10.1103/PhysRevD.105.034027}
  {\path{doi:10.1103/PhysRevD.105.034027}}.

\bibitem{DAlesio:2010sag}
U.~D'Alesio, F.~Murgia, C.~Pisano, {Azimuthal asymmetries for hadron
  distributions inside a jet in hadronic collisions}, Phys. Rev. D 83 (2011)
  034021.
\newblock \href {http://arxiv.org/abs/1011.2692} {\path{arXiv:1011.2692}},
  \href {https://doi.org/10.1103/PhysRevD.83.034021}
  {\path{doi:10.1103/PhysRevD.83.034021}}.

\bibitem{Kang:2017btw}
Z.-B. Kang, A.~Prokudin, F.~Ringer, F.~Yuan, {Collins azimuthal asymmetries of
  hadron production inside jets}, Phys. Lett. B 774 (2017) 635.
\newblock \href {http://arxiv.org/abs/1707.00913} {\path{arXiv:1707.00913}},
  \href {https://doi.org/10.1016/j.physletb.2017.10.031}
  {\path{doi:10.1016/j.physletb.2017.10.031}}.

\bibitem{Kang:2017glf}
Z.-B. Kang, X.~Liu, F.~Ringer, H.~Xing, {The transverse momentum distribution
  of hadrons within jets}, JHEP 11 (2017) 068.
\newblock \href {http://arxiv.org/abs/1705.08443} {\path{arXiv:1705.08443}},
  \href {https://doi.org/10.1007/JHEP11(2017)068}
  {\path{doi:10.1007/JHEP11(2017)068}}.

\bibitem{Bain:2016rrv}
R.~Bain, Y.~Makris, T.~Mehen, {Transverse Momentum Dependent Fragmenting Jet
  Functions with Applications to Quarkonium Production}, JHEP 11 (2016) 144.
\newblock \href {http://arxiv.org/abs/1610.06508} {\path{arXiv:1610.06508}},
  \href {https://doi.org/10.1007/JHEP11(2016)144}
  {\path{doi:10.1007/JHEP11(2016)144}}.

\bibitem{Kang:2020xyq}
Z.-B. Kang, K.~Lee, F.~Zhao, {Polarized jet fragmentation functions}, Phys.
  Lett. B 809 (2020) 135756.
\newblock \href {http://arxiv.org/abs/2005.02398} {\path{arXiv:2005.02398}},
  \href {https://doi.org/10.1016/j.physletb.2020.135756}
  {\path{doi:10.1016/j.physletb.2020.135756}}.

\bibitem{Kang:2023elg}
Z.-B. Kang, H.~Xing, F.~Zhao, Y.~Zhou {Polarized fragmenting jet functions in inclusive and exclusive jet production}, \newblock \href {http://arxiv.org/abs/2311.00672} {\path{arXiv:2311.00672}}.

\bibitem{Kang:2021ffh}
Z.-B. Kang, K.~Lee, D.~Y. Shao, F.~Zhao, {Spin asymmetries in electron-jet
  production at the future electron ion collider}, JHEP 11 (2021) 005.
\newblock \href {http://arxiv.org/abs/2106.15624} {\path{arXiv:2106.15624}},
  \href {https://doi.org/10.1007/JHEP11(2021)005}
  {\path{doi:10.1007/JHEP11(2021)005}}.

\bibitem{DAlesio:2017bvu}
U.~D'Alesio, F.~Murgia, C.~Pisano, {Testing the universality of the Collins
  function in pion-jet production at RHIC}, Phys. Lett. B 773 (2017) 300.
\newblock \href {http://arxiv.org/abs/1707.00914} {\path{arXiv:1707.00914}},
  \href {https://doi.org/10.1016/j.physletb.2017.08.023}
  {\path{doi:10.1016/j.physletb.2017.08.023}}.


\bibitem{Gao:2024dxl}
T.~Gao, et~al., STAR Collaboration,
{Measurement of transverse polarization of $\Lambda$/$\bar{\Lambda}$ within jet in $pp$ collisions at STAR}, Proceedings of the 25th International Spin Symposium (SPIN2023) (2023).
\newblock \href {http://arxiv.org/abs/2402.01168} {\path{arXiv:2402.01168}}

\bibitem{spinGao}
T.~Gao, et~al., STAR Collaboration,
 \href{https://indico.jlab.org/event/663/contributions/13260/}{Measurement of  transverse polarization of {$\Lambda$} in $pp$ collisions at {STAR}}, 25th  International Spin Symposium (SPIN2023) (2023).
\newline\urlprefix\url{https://indico.jlab.org/event/663/contributions/13260/}

\bibitem{Bacchetta:2004jz}
A.~Bacchetta, U.~D'Alesio, M.~Diehl, C.~A. Miller, Single-spin asymmetries: The
  {T}rento conventions, Phys. Rev. D 70 (2004) 117504.
\newblock \href {http://arxiv.org/abs/hep-ph/0410050}
  {\path{arXiv:hep-ph/0410050}}, \href
  {https://doi.org/10.1103/PhysRevD.70.117504}
  {\path{doi:10.1103/PhysRevD.70.117504}}.

\bibitem{DAlesio:2021dcx}
U.~D'Alesio, F.~Murgia, M.~Zaccheddu, {General helicity formalism for
  two-hadron production in $e^+e^-$ annihilation within a TMD approach}, JHEP
  10 (2021) 078.
\newblock \href {http://arxiv.org/abs/2108.05632} {\path{arXiv:2108.05632}},
  \href {https://doi.org/10.1007/JHEP10(2021)078}
  {\path{doi:10.1007/JHEP10(2021)078}}.

\bibitem{Cacciari:2008gp}
M.~Cacciari, G.~P. Salam, G.~Soyez, {The anti-$k_t$ jet clustering algorithm},
  JHEP 04 (2008) 063.
\newblock \href {http://arxiv.org/abs/0802.1189} {\path{arXiv:0802.1189}},
  \href {https://doi.org/10.1088/1126-6708/2008/04/063}
  {\path{doi:10.1088/1126-6708/2008/04/063}}.

\bibitem{Kaufmann:2015hma}
T.~Kaufmann, A.~Mukherjee, W.~Vogelsang, {Hadron Fragmentation Inside Jets in
  Hadronic Collisions}, Phys. Rev. D 92 (2015) 054015, [Erratum: Phys.Rev.D
  101, 079901 (2020)].
\newblock \href {http://arxiv.org/abs/1506.01415} {\path{arXiv:1506.01415}},
  \href {https://doi.org/10.1103/PhysRevD.92.054015}
  {\path{doi:10.1103/PhysRevD.92.054015}}.

\bibitem{Makris:2020ltr}
Y.~Makris, F.~Ringer, W.~J. Waalewijn, {Joint thrust and TMD resummation in
  electron-positron and electron-proton collisions}, JHEP 02 (2021) 070.
\newblock \href {http://arxiv.org/abs/2009.11871} {\path{arXiv:2009.11871}},
  \href {https://doi.org/10.1007/JHEP02(2021)070}
  {\path{doi:10.1007/JHEP02(2021)070}}.

\bibitem{Boglione:2020cwn}
M.~Boglione, A.~Simonelli, {Universality-breaking effects in $e^+e^-$ hadronic
  production processes}, Eur. Phys. J. C 81 (2021) 96.
\newblock \href {http://arxiv.org/abs/2007.13674} {\path{arXiv:2007.13674}},
  \href {https://doi.org/10.1140/epjc/s10052-020-08821-y}
  {\path{doi:10.1140/epjc/s10052-020-08821-y}}.

\bibitem{Boglione:2021wov}
M.~Boglione, A.~Simonelli, {Kinematic regions in the $e^+e^- \to h \, X$
  factorized cross section in a $2$-jet topology with thrust}, JHEP 02 (2022) 013.
\newblock \href {http://arxiv.org/abs/2109.11497} {\path{arXiv:2109.11497}},
  \href {https://doi.org/10.1007/JHEP02(2022)013}
  {\path{doi:10.1007/JHEP02(2022)013}}.

\bibitem{Boglione:2022nzq}
M.~Boglione, J.~O. Gonzalez-Hernandez, A.~Simonelli, {Transverse momentum
  dependent fragmentation functions from recent BELLE data}, Phys. Rev. D
  106 (2022) 074024.
\newblock \href {http://arxiv.org/abs/2206.08876} {\path{arXiv:2206.08876}},
  \href {https://doi.org/10.1103/PhysRevD.106.074024}
  {\path{doi:10.1103/PhysRevD.106.074024}}.

\bibitem{Dulat:2015mca}
S.~Dulat, T.-J. Hou, J.~Gao, M.~Guzzi, J.~Huston, P.~Nadolsky, J.~Pumplin,
  C.~Schmidt, D.~Stump, C.~P. Yuan, {New parton distribution functions from a
  global analysis of quantum chromodynamics}, Phys. Rev. D 93 (2016)
  033006.
\newblock \href {http://arxiv.org/abs/1506.07443} {\path{arXiv:1506.07443}},
  \href {https://doi.org/10.1103/PhysRevD.93.033006}
  {\path{doi:10.1103/PhysRevD.93.033006}}.

\bibitem{Boglione:2024dal}
M.~Boglione, U.~D'Alesio, C.~Flore, J.~O.~Gonzalez-Hernandez, F.~Murgia and A.~Prokudin,
{Simultaneous reweighting of Transverse Momentum Dependent distributions}, 
\newblock \href {http://arxiv.org/abs/2402.12322} {\path{arXiv:2402.12322}}.

\bibitem{AbdulKhalek:2021gbh}
R.~Abdul~Khalek, et~al., {Science Requirements and Detector Concepts for the
  Electron-Ion Collider}: {EIC Yellow Report}, Nucl. Phys. A 1026 (2022)
  122447.
\newblock \href {http://arxiv.org/abs/2103.05419} {\path{arXiv:2103.05419}},
  \href {https://doi.org/10.1016/j.nuclphysa.2022.122447}
  {\path{doi:10.1016/j.nuclphysa.2022.122447}}.

\bibitem{Aybat:2011zv}
S.~M. Aybat, T.~C. Rogers, {Transverse momentum dependent parton distribution
  and fragmentation functions with QCD evolution}, Phys. Rev. D 83 (2011)
  114042.
\newblock \href {http://arxiv.org/abs/1101.5057} {\path{arXiv:1101.5057}},
  \href {https://doi.org/10.1103/PhysRevD.83.114042}
  {\path{doi:10.1103/PhysRevD.83.114042}}.

\bibitem{Collins:2017oxh}
J.~Collins, T.~C. Rogers, {Connecting different TMD factorization formalisms in
  QCD}, Phys. Rev. D 96 (2017) 054011.
\newblock \href {http://arxiv.org/abs/1705.07167} {\path{arXiv:1705.07167}},
  \href {https://doi.org/10.1103/PhysRevD.96.054011}
  {\path{doi:10.1103/PhysRevD.96.054011}}.

\bibitem{Echevarria:2016scs}
M.~G. Echevarria, I.~Scimemi, A.~Vladimirov, {Unpolarized Transverse Momentum
  Dependent Parton Distribution and Fragmentation Functions at
  next-to-next-to-leading order}, JHEP 09 (2016) 004.
\newblock \href {http://arxiv.org/abs/1604.07869} {\path{arXiv:1604.07869}},
  \href {https://doi.org/10.1007/JHEP09(2016)004}
  {\path{doi:10.1007/JHEP09(2016)004}}.

\bibitem{Bacchetta:2017gcc}
A.~Bacchetta, F.~Delcarro, C.~Pisano, M.~Radici, A.~Signori, {Extraction of
  partonic transverse momentum distributions from semi-inclusive deep-inelastic
  scattering, Drell-Yan and Z-boson production}, JHEP 06 (2017) 081.
\newblock \href {http://arxiv.org/abs/1703.10157} {\path{arXiv:1703.10157}},
  \href {https://doi.org/10.1007/JHEP06(2017)081}
  {\path{doi:10.1007/JHEP06(2017)081}}.

\end{thebibliography}
\end{document}